# On the calculation rule of probability of relativistic free particle in quantum mechanics

T. Mei

(Department of Journal, Central China Normal University, Wuhan, Hubei PRO, People's Republic of China
E-Mail: meitao@mail.ccnu.edu.cn meitaowh@public.wh.hb.cn )

**Abstract**: As is well known, in quantum mechanics, the calculation rule of the probability that an eigen-value $a_n$ is observed when the physical quantity $A$ is measured for a state described by the state vector $|\Psi\rangle$ is $P(a_n) = \langle\Psi|A_n\rangle\langle A_n|\Psi\rangle$. However, in Ref.[1], based on strict logical reasoning and mathematical calculation, it has been pointed out, replacing $\langle\Psi|A_n\rangle\langle A_n|\Psi\rangle$, one should use a new rule to calculate $P(a_n)$ for particle satisfying the Dirac equation.

In this paper, we first state some results given by Ref.[1]. And then, we present a proof for the new calculation rule of probability according to Dirac sea of negative energy particles, hole theory and the principle "the vacuum is not observable". Finally, we discuss simply the case of particle satisfying the Klein-Gordon equation.

## 1 Some results given by Ref.[1]

### 1.1 Definition and determination of some distribution functions

Definitions of the three distribution functions $\Lambda(\boldsymbol{v})$, $f(\boldsymbol{p})$ and $\rho(t, \boldsymbol{x}; t_0, \boldsymbol{x}_0)$ are as follows:

① The velocity distribution function of a ponit particle $\Lambda(\boldsymbol{v})$: When the position of a point particle is fully measured at a moment, the probability that the value of the velocity of the point particle is situated between $\boldsymbol{v}$ and $\boldsymbol{v} + \mathrm{d}\boldsymbol{v}$ is $\Lambda(\boldsymbol{v})\mathrm{d}^3 v$.

② The momentum distribution function of a ponit particle $f(\boldsymbol{p})$: When the position of a point particle is fully measured at a moment, the probability that the value of the momentum of the point particle is situated between $\boldsymbol{p}$ and $\boldsymbol{p} + \mathrm{d}\boldsymbol{p}$ is $f(\boldsymbol{p})\mathrm{d}^3 p$.

③ The position distribution function of a free ponit particle $\rho(t, \boldsymbol{x}; t_0, \boldsymbol{x}_0)$: If a free point particle is situated at position $\boldsymbol{x}_0$ at time $t_0$, then the probability that the free point particle is situated between $\boldsymbol{x}$ and $\boldsymbol{x} + \mathrm{d}\boldsymbol{x}$ at time $t$ is $\rho(t, \boldsymbol{x}; t_0, \boldsymbol{x}_0)\mathrm{d}^3 x$, where $t > t_0$.

Based on the principle *the values of probability measured in different inertial frames are the same* and some mathematical derivation, the above three distribution functions are determinated, and the concrete forms of nonrelativistic and relativistic cases are given by Tab. 1 and Tab. 2, respectively.

.

**Tab. 1 Various distribution density functions in the nonrelativistic case**

| Physical quantity | | Distribution density function |
|---|---|---|
| velocity | | $\Lambda_1(\boldsymbol{v}) = \Lambda_1(0) = \text{constant}$ |
| momentum | | $f_1(\boldsymbol{p}) = \dfrac{1}{(2\pi)^3}$ |
| Free particle | position density | $\rho_1(t, \boldsymbol{x}; t_0, \boldsymbol{x}_0) = \dfrac{1}{(2\pi)^3} \dfrac{m^3}{(t - t_0)^3}$ |
| | current density | $\boldsymbol{j}_1(t, \boldsymbol{x}) = \dfrac{1}{(2\pi)^3} \dfrac{m^3(\boldsymbol{x} - \boldsymbol{x}_0)}{(t - t_0)^4}$ |

**Tab. 2 Various distribution density functions in the relativistic case**

| Physical quantity | | Distribution density function |
|---|---|---|
| velocity | | $\Lambda_2(\boldsymbol{v}) = \dfrac{\Lambda_2(0)}{(1 - v^2)^2}$ |
| momentum | | $f_2(\boldsymbol{p}) = \dfrac{1}{(2\pi)^3} \dfrac{1}{\sqrt{1 + p^2 / m^2}}$ |
| Free particle | position density | $\rho_2(t, \boldsymbol{x}; t_0, \boldsymbol{x}_0) = \begin{cases} \dfrac{1}{(2\pi)^3} \dfrac{m^3(t - t_0)}{\left[(t - t_0)^2 - (\boldsymbol{x} - \boldsymbol{x}_0)^2\right]^2}, & \lvert \boldsymbol{x} - \boldsymbol{x}_0 \rvert < t - t_0 ; \\ 0, & \lvert \boldsymbol{x} - \boldsymbol{x}_0 \rvert \geq t - t_0 . \end{cases}$ |
| | current density | $\boldsymbol{j}_2(t, \boldsymbol{x}) = \begin{cases} \dfrac{1}{(2\pi)^3} \dfrac{m^3(\boldsymbol{x} - \boldsymbol{x}_0)}{\left[(t - t_0)^2 - (\boldsymbol{x} - \boldsymbol{x}_0)^2\right]^2}, & \lvert \boldsymbol{x} - \boldsymbol{x}_0 \rvert < t - t_0 ; \\ 0, & \lvert \boldsymbol{x} - \boldsymbol{x}_0 \rvert \geq t - t_0 . \end{cases}$ |

In Tab. 1 and Tab. 2, the form of the current density function of a free point particle $\boldsymbol{j}(t, \boldsymbol{x})$ is given by

$$\boldsymbol{j}_i(t, \boldsymbol{x}) = \rho_i(t, \boldsymbol{x}; t_0, \boldsymbol{x}_0)\boldsymbol{v} = \rho_i(t, \boldsymbol{x}; t_0, \boldsymbol{x}_0) \frac{\boldsymbol{x} - \boldsymbol{x}_0}{t - t_0} \quad (i = 1, 2),$$

and from Tab. 1 and Tab. 2 we can verify easily the current conservation

$$\frac{\partial \rho_i(t, \boldsymbol{x}; t_0, \boldsymbol{x}_0)}{\partial t} + \nabla \cdot \boldsymbol{j}_i(t, \boldsymbol{x}) = 0 \quad (i = 1, 2).$$

**1.2 Comparing the corresponding conclusions in quantum mechanics with the results in Tab. 1 and Tab.2**

In Ref. [1], it is point out that the conclusions about nonrelativistic particle in quantum mechanics are in accord with Tab. 1, but the conclusions about relativistic particle in quantum mechanics are not in accord with Tab. 2.

At first, in quantum mechanics, when a point particle is situated at position $\boldsymbol{x}_0$ at time $t_0$, the state of the point particle is described by the state vector $|t_0\rangle = |\boldsymbol{x}_0\rangle$; and, no matter whether it is nonrelativistic or relativistic case, if we measure momentum for this state, then the probability density $f(\boldsymbol{p})$ that momentum volue $\boldsymbol{p}$ is measured is

$$f(\boldsymbol{p}) = \langle \boldsymbol{x}_0 | \boldsymbol{p} \rangle \langle \boldsymbol{p} | \boldsymbol{x}_0 \rangle = \frac{1}{(2\pi)^{3/2}} \mathrm{e}^{-\mathrm{i}\boldsymbol{p}\cdot\boldsymbol{x}_0} \cdot \frac{1}{(2\pi)^{3/2}} \mathrm{e}^{\mathrm{i}\boldsymbol{p}\cdot\boldsymbol{x}_0} = \frac{1}{(2\pi)^3}.$$

Comparing $f(\boldsymbol{p})$ with $f_1(\boldsymbol{p})$ in Tab. 1 and $f_2(\boldsymbol{p})$ in Tab. 2, respectively, we see that $f(\boldsymbol{p}) = f_1(\boldsymbol{p})$ in nonrelativistic case but $f(\boldsymbol{p}) \neq f_2(\boldsymbol{p})$ in relativistic case.

Secondly, in quantum mechanics, the physical meaning of the propagator $K(t, \boldsymbol{x}; t_0, \boldsymbol{x}_0)$ of a particle is

$$K^+(t, \boldsymbol{x}; t_0, \boldsymbol{x}_0) \cdot K(t, \boldsymbol{x}; t_0, \boldsymbol{x}_0) = \rho(t, \boldsymbol{x}; t_0, \boldsymbol{x}_0).$$

And, further, the propagator of a nonrelativistic free particle is

$$K_0(t, \boldsymbol{x}; t_0, \boldsymbol{x}_0) = \int \frac{\mathrm{d}^3 p}{(2\pi)^3} \mathrm{e}^{\mathrm{i}\boldsymbol{p}\cdot(\boldsymbol{x}-\boldsymbol{x}_0)} \mathrm{e}^{-\mathrm{i}\frac{p^2}{2m}(t-t_0)} = \left[\frac{m}{2\pi\mathrm{i}(t-t_0)}\right]^{\frac{3}{2}} \mathrm{e}^{\mathrm{i}\frac{(\boldsymbol{x}-\boldsymbol{x}_0)^2}{t-t_0}},$$

we therefore have

$$K_0^+(t, \boldsymbol{x}; t_0, \boldsymbol{x}_0) \cdot K_0(t, \boldsymbol{x}; t_0, \boldsymbol{x}_0) = \frac{1}{(2\pi)^3} \frac{m^3}{(t-t_0)^3} = \rho_1(t, \boldsymbol{x}; t_0, \boldsymbol{x}_0),$$

where $\rho_1(t, \boldsymbol{x}; t_0, \boldsymbol{x}_0)$ is given by Tab. 1. We see that such result is in accord with Tab. 1.

On the other hand, in relativistic case, the propagator $K_\mathrm{D}(t, \boldsymbol{x}, t_0, \boldsymbol{x}_0)$ of a free particle determined by the Dirac Hamiltonian $H_\mathrm{D} = \boldsymbol{\alpha} \cdot \hat{\boldsymbol{p}} + \beta m$ is[2]

$$\begin{aligned}
K_\mathrm{D}(t, \boldsymbol{x}, t_0, \boldsymbol{x}_0) &= \int \frac{\mathrm{d}^3 p}{(2\pi)^3} \mathrm{e}^{\mathrm{i}\boldsymbol{p}\cdot(\boldsymbol{x}-\boldsymbol{x}_0)} \mathrm{e}^{-\mathrm{i}(\boldsymbol{\alpha}\cdot\boldsymbol{p}+\beta m)(t-t_0)} \\
&= \int \frac{\mathrm{d}^3 p}{(2\pi)^3} \mathrm{e}^{\mathrm{i}\boldsymbol{p}\cdot(\boldsymbol{x}-\boldsymbol{x}_0)} \left\{ \cos\left[\sqrt{p^2+m^2}(t-t_0)\right] - \mathrm{i}\frac{\boldsymbol{\alpha}\cdot\boldsymbol{p}+\beta m}{\sqrt{p^2+m^2}} \sin\left[\sqrt{p^2+m^2}(t-t_0)\right] \right\} \\
&= \mathrm{i}\left(\mathrm{i}\frac{\partial}{\partial t} + \boldsymbol{\alpha}\cdot\frac{\partial}{\mathrm{i}\partial \boldsymbol{x}} + \beta \mathrm{m}\right) \Delta(x - x_0),
\end{aligned}$$

where[2]

$$\Delta(x) = -\int \frac{\mathrm{d}^3 p}{(2\pi)^3} \mathrm{e}^{\mathrm{i}\boldsymbol{p}\cdot\boldsymbol{x}} \frac{\sin\left(\sqrt{p^2+m^2 t}\right)}{\sqrt{p^2+m^2}} = \frac{1}{4\pi r}\frac{\partial}{\partial r} \begin{cases} J_0(ms), & |\boldsymbol{x}| < t; \\ 0, & |\boldsymbol{x}| > t. \end{cases} \tag{1}$$

In the above formula, $J_0(ms)$ is a Bseeel function, $s = \sqrt{t^2 - \boldsymbol{x}^2}$. Hence, when $|\boldsymbol{x}| > t$,

$$K_D^+(t,\boldsymbol{x};t_0,\boldsymbol{x}_0)\cdot K_D(t,\boldsymbol{x};t_0,\boldsymbol{x}_0)=0\,;$$

however, when $|\boldsymbol{x}|<t$, from (1) we can be prove

$$K_D^+(t,\boldsymbol{x};t_0,\boldsymbol{x}_0)\cdot K_D(t,\boldsymbol{x};t_0,\boldsymbol{x}_0)\neq\frac{1}{(2\pi)^3}\frac{m^3(t-t_0)}{\left[(t-t_0)^2-(\boldsymbol{x}-\boldsymbol{x}_0)^2\right]^2},$$

hence,

$$K_D^+(t,\boldsymbol{x};t_0,\boldsymbol{x}_0)\cdot K_D(t,\boldsymbol{x};t_0,\boldsymbol{x}_0)\neq\rho_2(t,\boldsymbol{x};t_0,\boldsymbol{x}_0),$$

where $\rho_2(t,\boldsymbol{x};t_0,\boldsymbol{x}_0)$ is given by Tab. 2.

### 1.3 Three mathematical formulas

The following three mathematical formulas have been given and proved in Ref.[1].

Defining a function $I_1(t,\boldsymbol{x})$:

$$I_1(t,\boldsymbol{x})\equiv\int\frac{\mathrm{d}^3p}{(2\pi)^3}\mathrm{e}^{\mathrm{i}\boldsymbol{p}\cdot\boldsymbol{x}}\cos\left(\sqrt{p^2+m^2}\,t\right)\times\int\frac{\mathrm{d}^3p'}{(2\pi)^3}\mathrm{e}^{\mathrm{i}\boldsymbol{p}'\cdot\boldsymbol{x}}\frac{m}{\sqrt{p'^2+m^2}}\cos\left(\sqrt{p'^2+m^2}\,t\right)$$
$$+\int\frac{\mathrm{d}^3p}{(2\pi)^3}\mathrm{e}^{\mathrm{i}\boldsymbol{p}\cdot\boldsymbol{x}}\sin\left(\sqrt{p^2+m^2}\,t\right)\times\int\frac{\mathrm{d}^3p'}{(2\pi)^3}\mathrm{e}^{\mathrm{i}\boldsymbol{p}'\cdot\boldsymbol{x}}\frac{m}{\sqrt{p'^2+m^2}}\sin\left(\sqrt{p'^2+m^2}\,t\right),$$

**the first formula** is

$$I_1(t,\boldsymbol{x})=\begin{cases}\dfrac{1}{(2\pi)^3}\dfrac{m^3t}{\left(t^2-\boldsymbol{x}^2\right)^2}, & |\boldsymbol{x}|<t\,;\\ 0, & |\boldsymbol{x}|\geq t\,.\end{cases}$$

Notice that

$$I_1(t-\boldsymbol{x},t_0-\boldsymbol{x}_0)=\rho_2(t,\boldsymbol{x};t_0,\boldsymbol{x}_0)\,.$$

And then, based on the following operators and functions

$$\hat{M}_\pm=\sum_{\boldsymbol{p}}|\boldsymbol{p}\rangle\frac{1}{2}\left(1\pm\frac{\boldsymbol{\alpha}\cdot\boldsymbol{p}+\beta m}{\sqrt{p^2+m^2}}\right)\langle\boldsymbol{p}|,\quad M_\pm(\boldsymbol{x},\boldsymbol{x}')=\int\frac{\mathrm{d}^3p}{(2\pi)^3}\mathrm{e}^{\mathrm{i}\boldsymbol{p}\cdot(\boldsymbol{x}-\boldsymbol{x}')}\frac{1}{2}\left(1\pm\frac{\boldsymbol{\alpha}\cdot\boldsymbol{p}+\beta m}{\sqrt{p^2+m^2}}\right).$$

$$K_{D\pm}(t,\boldsymbol{x};t_0,\boldsymbol{x}_0)=\int M_\pm(\boldsymbol{x},\boldsymbol{x}')\mathrm{d}^3x'K_D(t,\boldsymbol{x}';t_0,\boldsymbol{x}_0),$$

we consider a function $I_2(t,\boldsymbol{x};t_0,\boldsymbol{x}_0)$:

$$I_2(t,\boldsymbol{x};t_0,\boldsymbol{x}_0)\equiv K_{D+}^+(t,\boldsymbol{x};t_0,\boldsymbol{x}_0)\cdot K_{D+}(t,\boldsymbol{x};t_0,\boldsymbol{x}_0)$$
$$-K_{D-}^+(t,\boldsymbol{x};t_0,\boldsymbol{x}_0)\cdot K_{D-}(t,\boldsymbol{x};t_0,\boldsymbol{x}_0),$$

**the second formula** is

$$I_2(t,\boldsymbol{x};t_0,\boldsymbol{x}_0)=\beta I_1(t-\boldsymbol{x},t_0-\boldsymbol{x}_0)=\beta\rho_2(t,\boldsymbol{x};t_0,\boldsymbol{x}_0)$$
$$=\begin{cases}\beta\dfrac{1}{(2\pi)^3}\dfrac{m^3(t-t_0)}{\left[(t-t_0)^2-(\boldsymbol{x}-\boldsymbol{x}_0)^2\right]^2}, & |\boldsymbol{x}-\boldsymbol{x}_0|<t-t_0\,;\\ 0, & |\boldsymbol{x}-\boldsymbol{x}_0|\geq t-t_0\,.\end{cases}$$

Finally, for a function $\boldsymbol{I}_3(t,\boldsymbol{x};t_0,\boldsymbol{x}_0)$:

$$\boldsymbol{I}_3(t,\boldsymbol{x};t_0,\boldsymbol{x}_0)\equiv K_{\mathrm{D}+}^{+}(t,\boldsymbol{x};t_0,\boldsymbol{x}_0)\cdot\boldsymbol{\alpha}\cdot K_{\mathrm{D}+}(t,\boldsymbol{x};t_0,\boldsymbol{x}_0)$$
$$-K_{\mathrm{D}-}^{+}(t,\boldsymbol{x};t_0,\boldsymbol{x}_0)\cdot\boldsymbol{\alpha}\cdot K_{\mathrm{D}-}(t,\boldsymbol{x};t_0,\boldsymbol{x}_0),$$

**the third formula** is

$$\boldsymbol{I}_3(t,\boldsymbol{x};t_0,\boldsymbol{x}_0)=\beta\,\boldsymbol{j}_2(t,\boldsymbol{x})$$
$$=\begin{cases}\beta\dfrac{1}{(2\pi)^3}\dfrac{m^3(\boldsymbol{x}-\boldsymbol{x}_0)}{\left[(t-t_0)^2-(\boldsymbol{x}-\boldsymbol{x}_0)^2\right]^2}, & |\boldsymbol{x}-\boldsymbol{x}_0|<t-t_0\,;\\ 0, & |\boldsymbol{x}-\boldsymbol{x}_0|\geq t-t_0\,.\end{cases}$$

where $\boldsymbol{j}_2(t,\boldsymbol{x})$ is given by Tab. 2.

### 1.4 New calculation rule of probability

Based on the above discussion, in Ref.[1], a new calculation rule of probability for a free particle satisfying the Dirac equation has been given as follows.

For an arbitrary state described by the state vector $|\widetilde{\Psi}\rangle$, we define

$$|\widetilde{\Psi}_{\pm}\rangle=\hat{M}_{\pm}|\widetilde{\Psi}\rangle\,;$$

For a physical quantity $A$ of which the eigenvalue equation reads

$$\hat{A}A_n(\boldsymbol{x})=a_nA_n(\boldsymbol{x}), \tag{2}$$

the probability $P(a_n)$ that an eigen-value $a_n$ is observed when the physical quantity $A$ is measured for the state $|\widetilde{\Psi}\rangle$ is:

$$\begin{aligned}P(a_n)&=\langle\widetilde{\Psi}_+|A_n\rangle\langle A_n|\widetilde{\Psi}_+\rangle-\langle\widetilde{\Psi}_-|A_n\rangle\langle A_n|\widetilde{\Psi}_-\rangle\\&=\langle\widetilde{\Psi}|\hat{M}_+|A_n\rangle\langle A_n|\hat{M}_+|\widetilde{\Psi}\rangle-\langle\widetilde{\Psi}|\hat{M}_-|A_n\rangle\langle A_n|\hat{M}_-|\widetilde{\Psi}\rangle\,.\end{aligned} \tag{3}$$

In Ref.[1], it has been proved that, the results obtained by the above calculation rule (3) are in accord with Tab. 2 for a free particle satisfying the Dirac equation. Especially, taking advantage of the above thirt formula, it has been proved that the three dimensioal current density $\langle\widetilde{\Psi}_+|\boldsymbol{x}\rangle\boldsymbol{\alpha}\langle\boldsymbol{x}|\widetilde{\Psi}_+\rangle-\langle\widetilde{\Psi}_-|\boldsymbol{x}\rangle\boldsymbol{\alpha}\langle\boldsymbol{x}|\widetilde{\Psi}_-\rangle$ corresponding to position probability density $\langle\widetilde{\Psi}_+|\boldsymbol{x}\rangle\langle\boldsymbol{x}|\widetilde{\Psi}_+\rangle-\langle\widetilde{\Psi}_-|\boldsymbol{x}\rangle\langle\boldsymbol{x}|\widetilde{\Psi}_-\rangle$ obtained by (3) for a free particle equals $\boldsymbol{j}_2(t,\boldsymbol{x})$ in Tab. 2, hence, so called "zitterbewegung"[2] does not appear.

In this paper, we prove that the calculation rule of probability (3) can be derived from the idea of Dirac sea of negative energy particles, hole theory and the principle "*the vacuum is not observable*".

## 2 Discussion in coordinate representation

### 2.1 Some well known characteristics of the Dirac theory

We considering the complete set of eigenstates $\left\{ |U_r(\boldsymbol{p})\rangle, |V_r(\boldsymbol{p})\rangle, r = -\frac{1}{2}, \frac{1}{2} \right\}$ of the Dirac Hamilton $H_\mathrm{D} = \boldsymbol{\alpha} \cdot \hat{\boldsymbol{p}} + \beta m$, which satisfy

$$(\boldsymbol{\alpha} \cdot \hat{\boldsymbol{p}} + \beta m)|U_r(\boldsymbol{p})\rangle = \sqrt{p^2 + m^2}|U_r(\boldsymbol{p})\rangle,\ (\boldsymbol{\alpha} \cdot \hat{\boldsymbol{p}} + \beta m)|V_r(\boldsymbol{p})\rangle = -\sqrt{p^2 + m^2}|V_r(\boldsymbol{p})\rangle, \tag{4}$$

$$\langle U_r(\boldsymbol{p})|U_s(\boldsymbol{p})\rangle = \delta_{rs},\ \langle V_r(\boldsymbol{p})|V_s(\boldsymbol{p})\rangle = \delta_{rs},\ \langle U_r(\boldsymbol{p})|V_s(\boldsymbol{p})\rangle = 0,\ \langle V_r(\boldsymbol{p})|U_s(\boldsymbol{p})\rangle = 0, \tag{5}$$

$$\sum_{r=1}^{2}|U_r(\boldsymbol{p})\rangle\langle U_r(\boldsymbol{p})| = \frac{1}{2}\left(1 + \frac{\boldsymbol{\alpha} \cdot \hat{\boldsymbol{p}} + \beta m}{\sqrt{p^2 + m^2}}\right),\ \sum_{r=1}^{2}|V_r(\boldsymbol{p})\rangle\langle V_r(\boldsymbol{p})| = \frac{1}{2}\left(1 - \frac{\boldsymbol{\alpha} \cdot \hat{\boldsymbol{p}} + \beta m}{\sqrt{p^2 + m^2}}\right). \tag{6}$$

The corresponding wave functions $U_r(\boldsymbol{p}, \boldsymbol{x})$ $(r = 1, 2)$ and $V_r(\boldsymbol{p}, \boldsymbol{x})$ $(r = 1, 2)$ are:

$$U_r(\boldsymbol{p}, \boldsymbol{x}) = \langle \boldsymbol{x}|U_r(\boldsymbol{p})\rangle = U_r(\boldsymbol{p})\frac{\mathrm{e}^{\mathrm{i}\boldsymbol{p}\cdot\boldsymbol{x}}}{(2\pi)^{3/2}},\ V_r(\boldsymbol{p}, \boldsymbol{x}) = \langle \boldsymbol{x}|V_r(\boldsymbol{p})\rangle = V_r(\boldsymbol{p})\frac{\mathrm{e}^{\mathrm{i}\boldsymbol{p}\cdot\boldsymbol{x}}}{(2\pi)^{3/2}}. \tag{7}$$

The wave function $\widetilde{\Psi}(t, \boldsymbol{x})$ of a state $|\widetilde{\Psi}(t)\rangle$ can be written to the form

$$\widetilde{\Psi}(t, \boldsymbol{x}) = \langle \boldsymbol{x}|\widetilde{\Psi}\rangle = \int \mathrm{d}^3 p \sum_{r=1}^{2}\left(U_r(\boldsymbol{p}, \boldsymbol{x})A_r(t, \boldsymbol{p}) + V_r(\boldsymbol{p}, \boldsymbol{x})B_r(t, \boldsymbol{p})\right), \tag{8}$$

where $|\widetilde{\Psi}(t)\rangle$ describes a "one-particle" state (In fact, it should be replaced by a state vector of many particles system after considering Dirac sea, see the discussion of (14) below); and

$$\begin{aligned} A_r(t, \boldsymbol{p}) &= \langle U_r(\boldsymbol{p})|\widetilde{\Psi}(t)\rangle = \int \mathrm{d}^3 x U_r^+(\boldsymbol{p}, \boldsymbol{x})\Psi(t, \boldsymbol{x}), \\ B_r(t, \boldsymbol{p}) &= \langle V_r(\boldsymbol{p})|\widetilde{\Psi}(t)\rangle = \int \mathrm{d}^3 x V_r^+(\boldsymbol{p}, \boldsymbol{x})\Psi(t, \boldsymbol{x}), \end{aligned} \tag{9}$$

$$\langle \widetilde{\Psi}(t)|\widetilde{\Psi}(t)\rangle = \int \mathrm{d}^3 x \widetilde{\Psi}^+(t, \boldsymbol{x})\widetilde{\Psi}(t, \boldsymbol{x}) = \int \mathrm{d}^3 p \sum_{r=1}^{2}\left(A_r^*(\boldsymbol{p}, \boldsymbol{x})A_r(t, \boldsymbol{p}) + B_r^*(\boldsymbol{p}, \boldsymbol{x})B_r(t, \boldsymbol{p})\right) = 1. \tag{10}$$

### 2.2 The wave function of the vacuum when Dirac sea is as the vacuum

We now assume that the momentum of particle take quantized discrete values.

According to the model of Dirac sea and the theory of many particles system, we regard the vacuum as the state that the whole negative energy states are filled up, and the wave function of such vacuum $\Psi_\text{vacuum}(\boldsymbol{x}_1, \boldsymbol{x}_2, \cdots, \boldsymbol{x}_i, \cdots)$ is:

$$
\begin{aligned}
&\Psi_{\text{vacuum}}(\boldsymbol{x}_1,\boldsymbol{x}_2,\cdots,\boldsymbol{x}_i,\cdots)\\
&=\frac{1}{\sqrt{N!}}\Bigg(\sum_{(\boldsymbol{x}_{i_1},\boldsymbol{x}_{i_2},\cdots,\boldsymbol{x}_{i_{s'}},\cdots)}(-1)^{\delta(\boldsymbol{x}_{i_1},\boldsymbol{x}_{i_2},\cdots,\boldsymbol{x}_{i_{s'}},\cdots)}V_{-\frac{1}{2}}(\boldsymbol{p}_1,\boldsymbol{x}_{i_1})V_{\frac{1}{2}}(\boldsymbol{p}_1,\boldsymbol{x}_{i_2})V_{-\frac{1}{2}}(\boldsymbol{p}_2,\boldsymbol{x}_{i_3})V_{\frac{1}{2}}(\boldsymbol{p}_2,\boldsymbol{x}_{i_4})\cdots\\
&\quad\times V_{-\frac{1}{2}}(\boldsymbol{p}_{s-1},\boldsymbol{x}_{i_{2(s-1)-1}})V_{\frac{1}{2}}(\boldsymbol{p}_{s-1},\boldsymbol{x}_{i_{2(s-1)}})V_{-\frac{1}{2}}(\boldsymbol{p}_s,\boldsymbol{x}_{i_{2s-1}})V_{\frac{1}{2}}(\boldsymbol{p}_s,\boldsymbol{x}_{i_{2s}})\\
&\quad\times V_{-\frac{1}{2}}(\boldsymbol{p}_{s+1},\boldsymbol{x}_{i_{2(s+1)-1}})V_{\frac{1}{2}}(\boldsymbol{p}_{s+1},\boldsymbol{x}_{i_{2(s+1)}})\cdots\Bigg),
\end{aligned}
\tag{11}
$$

where the elements in the set $(\boldsymbol{p}_1,\boldsymbol{p}_2,\cdots,\boldsymbol{p}_s,\cdots)$ traverse the whole values of momentums, $N$ is the total number of the negative energy particles, $(\boldsymbol{x}_{i_1},\boldsymbol{x}_{i_2},\cdots,\boldsymbol{x}_{i_{s'}},\cdots)$ is a permutation of $(\boldsymbol{x}_1,\boldsymbol{x}_2,\cdots,\boldsymbol{x}_{s'},\cdots)$, the sum $\sum_{(\boldsymbol{x}_{i_1},\boldsymbol{x}_{i_2},\cdots,\boldsymbol{x}_{i_{s'}},\cdots)}$ traverse all permutations,

$$
\begin{aligned}
&\delta(\boldsymbol{x}_{i_1},\boldsymbol{x}_{i_2},\cdots,\boldsymbol{x}_{i_{s'}},\cdots)\\
&=\begin{cases}1, & \text{when } (\boldsymbol{x}_{i_1},\boldsymbol{x}_{i_2},\cdots,\boldsymbol{x}_{i_{s'}},\cdots) \text{ is even permutation of } (\boldsymbol{x}_1,\boldsymbol{x}_2,\cdots,\boldsymbol{x}_{s'},\cdots);\\ -1, & \text{when } (\boldsymbol{x}_{i_1},\boldsymbol{x}_{i_2},\cdots,\boldsymbol{x}_{i_{s'}},\cdots) \text{ is odd permutation of } (\boldsymbol{x}_1,\boldsymbol{x}_2,\cdots,\boldsymbol{x}_{s'},\cdots).\end{cases}
\end{aligned}
$$

### 2.3 The wave function of a particle when Dirac sea is as the vacuum

At first, for a particle situated positive energy state $|U_r(\boldsymbol{p})\rangle$, because there is now Dirac sea, we must use the wave function

$$
\begin{aligned}
&\Psi_r^{(+)}(\boldsymbol{p};\boldsymbol{p}_1,\boldsymbol{p}_2,\cdots,\boldsymbol{p}_{i'},\cdots;\boldsymbol{x}_1,\boldsymbol{x}_2,\cdots,\boldsymbol{x}_i,\cdots)\\
&=\frac{1}{\sqrt{N!}}\Bigg(\sum_{(\boldsymbol{x}_{i_1},\boldsymbol{x}_{i_2},\cdots,\boldsymbol{x}_{i_{s'}},\cdots)}(-1)^{\delta(\boldsymbol{x}_{i_1},\boldsymbol{x}_{i_2},\cdots,\boldsymbol{x}_{i_{s'}},\cdots)}U_r(\boldsymbol{p}_1,\boldsymbol{x}_{i_1})V_{-\frac{1}{2}}(\boldsymbol{p}_1,\boldsymbol{x}_{i_2})V_{\frac{1}{2}}(\boldsymbol{p}_1,\boldsymbol{x}_{i_3})\\
&\quad\times V_{-\frac{1}{2}}(\boldsymbol{p}_2,\boldsymbol{x}_{i_4})V_{\frac{1}{2}}(\boldsymbol{p}_2,\boldsymbol{x}_{i_5})\cdots V_{-\frac{1}{2}}(\boldsymbol{p}_{s-1},\boldsymbol{x}_{i_{2(s-1)}})V_{\frac{1}{2}}(\boldsymbol{p}_{s-1},\boldsymbol{x}_{i_{2(s-1)+1}})V_{-\frac{1}{2}}(\boldsymbol{p}_s,\boldsymbol{x}_{i_{2s}})V_{\frac{1}{2}}(\boldsymbol{p}_s,\boldsymbol{x}_{i_{2s+1}})\\
&\quad\times V_{-\frac{1}{2}}(\boldsymbol{p}_{s+1},\boldsymbol{x}_{i_{2(s+1)}})V_{\frac{1}{2}}(\boldsymbol{p}_{s+1},\boldsymbol{x}_{i_{2(s+1)+1}})\cdots\Bigg)
\end{aligned}
\tag{12}
$$

to describe such system in which there is a positive energy particle and Dirac sea.

Secondly, according to the Dirac hold theory, Dirac sea can be observed only when it lacks some negative energy state particles, hence, we can track formally the change of negative energy state by observing the motion of the hold (i.e., the motion of antiparticle). The concrete method has been given by Ref. [3]: If Dirac sea changes from the state lacking $|V_r(\boldsymbol{p})\rangle$ at time $t$ to another state lacking $|V_{r'}(\boldsymbol{p}')\rangle$ at time $t$+d$t$, then, formally, we can regard this process as that a negative energy particle moves from the state $|V_{r'}(\boldsymbol{p}')\rangle$ at time $t$+d$t$ to the state $|V_r(\boldsymbol{p})\rangle$ at time $t$. Hence, when we are tracking formally a negative energy state $|V_r(\boldsymbol{p})\rangle$, what we are actually observing is a many particles system described by the wave function

$$\Psi_r^{(-)}(\boldsymbol{p};\boldsymbol{p}_1,\boldsymbol{p}_2,\cdots,\boldsymbol{p}_{i'},\cdots;\boldsymbol{x}_1,\boldsymbol{x}_2,\cdots,\boldsymbol{x}_i,\cdots)$$

$$=(-1)^{\delta(r,\boldsymbol{p})+1}\frac{1}{\sqrt{N!}}\left(\sum_{(\boldsymbol{x}_{i_1},\boldsymbol{x}_{i_2},\cdots,\boldsymbol{x}_{i_{s'}},\cdots)}(-1)^{\delta(\boldsymbol{x}_{i_1},\boldsymbol{x}_{i_2},\cdots,\boldsymbol{x}_{i_{s'}},\cdots)}V_{-\frac{1}{2}}(\boldsymbol{p}_1,\boldsymbol{x}_{i_1})V_{\frac{1}{2}}(\boldsymbol{p}_1,\boldsymbol{x}_{i_2})V_{-\frac{1}{2}}(\boldsymbol{p}_2,\boldsymbol{x}_{i_3})V_{\frac{1}{2}}(\boldsymbol{p}_2,\boldsymbol{x}_{i_4})\cdots\right.$$

$$\left.\times V_{-\frac{1}{2}}(\boldsymbol{p}_{s-1},\boldsymbol{x}_{i_{2(s-1)-1}})V_{\frac{1}{2}}(\boldsymbol{p}_{s-1},\boldsymbol{x}_{i_{2(s-1)}})V_{-r}(\boldsymbol{p},\boldsymbol{x}_{i_{2s-1}})V_{-\frac{1}{2}}(\boldsymbol{p}_{s+1},\boldsymbol{x}_{i_{2s}})V_{\frac{1}{2}}(\boldsymbol{p}_{s+1},\boldsymbol{x}_{i_{2s+1}})\cdots\right); \tag{13}$$

The above expression lacks the term $V_r(\boldsymbol{p},\boldsymbol{x}_{i_{s''}})$ (we assume $\boldsymbol{p}=\boldsymbol{p}_s$), $\delta(r,\boldsymbol{p})$ is the "situation number" of $V_r(\boldsymbol{p},\boldsymbol{x}_{i_{s''}})$ situated in the expression

$$V_{-\frac{1}{2}}(\boldsymbol{p}_1,\boldsymbol{x}_{i_1})V_{\frac{1}{2}}(\boldsymbol{p}_1,\boldsymbol{x}_{i_2})V_{-\frac{1}{2}}(\boldsymbol{p}_2,\boldsymbol{x}_{i_3})V_{\frac{1}{2}}(\boldsymbol{p}_2,\boldsymbol{x}_{i_4})\cdots$$
$$\times V_{-\frac{1}{2}}(\boldsymbol{p}_{s-1},\boldsymbol{x}_{i_{2(s-1)-1}})V_{\frac{1}{2}}(\boldsymbol{p}_{s-1},\boldsymbol{x}_{i_{2(s-1)}})V_{-\frac{1}{2}}(\boldsymbol{p},\boldsymbol{x}_{i_{2s-1}})V_{\frac{1}{2}}(\boldsymbol{p},\boldsymbol{x}_{i_{2s}})$$
$$\times V_{-\frac{1}{2}}(\boldsymbol{p}_{s+1},\boldsymbol{x}_{i_{2(s+1)-1}})V_{\frac{1}{2}}(\boldsymbol{p}_{s+1},\boldsymbol{x}_{i_{2(s+1)}})\cdots,$$

namely, $V_r(\boldsymbol{p},\boldsymbol{x}_{i_{s''}})$ is situated on $\delta(r,\boldsymbol{p})$-th situation in the above expression.

Based on the above discussion, for a "one-particle" system described by the wave function (8), we must replace it by a wave function $\Psi(t;\boldsymbol{x}_1,\boldsymbol{x}_2,\cdots,\boldsymbol{x}_i,\cdots)$ of many particles system

$$\Psi(t;\boldsymbol{x}_1,\boldsymbol{x}_2,\cdots,\boldsymbol{x}_i,\cdots)=\Psi^{(1)}(t;\boldsymbol{x}_1,\boldsymbol{x}_2,\cdots,\boldsymbol{x}_i,\cdots)+\Psi^{(2)}(t;\boldsymbol{x}_1,\boldsymbol{x}_2,\cdots,\boldsymbol{x}_i,\cdots) \tag{14}$$

to describe this system, where

$$\Psi^{(1)}(t;\boldsymbol{x}_1,\boldsymbol{x}_2,\cdots,\boldsymbol{x}_i,\cdots)=\int \mathrm{d}^3p\sum_{r=1}^{2}A_r(t,\boldsymbol{p})\Psi_r^{(+)}(\boldsymbol{p};\boldsymbol{p}_1,\boldsymbol{p}_2,\cdots,\boldsymbol{p}_{i'},\cdots;\boldsymbol{x}_1,\boldsymbol{x}_2,\cdots,\boldsymbol{x}_i,\cdots), \tag{15}$$

$$\Psi^{(2)}(t;\boldsymbol{x}_1,\boldsymbol{x}_2,\cdots,\boldsymbol{x}_i,\cdots)=\int \mathrm{d}^3p\sum_{r=1}^{2}B_r^*(t,\boldsymbol{p})\Psi_r^{(-)}(\boldsymbol{p};\boldsymbol{p}_1,\boldsymbol{p}_2,\cdots,\boldsymbol{p}_{i'},\cdots;\boldsymbol{x}_1,\boldsymbol{x}_2,\cdots,\boldsymbol{x}_i,\cdots). \tag{16}$$

We emphasize that we must use $B_r^*(t,\boldsymbol{p})$ but not $B_r(t,\boldsymbol{p})$ in (16), or we cannot obtain correct result (See the calculation process in (44) below); on the other hand, this process is also in accord with the Feynman's theory of positrons[3], we explain this point in slightly detail.

We use $|\widetilde{\Psi}_r(\boldsymbol{p})\rangle_{\text{electron}}$ and $|\widetilde{\Psi}_r(\boldsymbol{p})\rangle_{\text{positron}}$ denote the state of electron and positron with momentum $p$ and spin $r$, respectively; if the initial and final state of a electron are $|\widetilde{\Psi}_i\rangle_{\text{electron}}=|\widetilde{\Psi}_r(\boldsymbol{p})\rangle_{\text{electron}}$ and $|\widetilde{\Psi}_f\rangle_{\text{electron}}=|\widetilde{\Psi}_{r'}(\boldsymbol{p}')\rangle_{\text{electron}}$, respectively, then the transition amplitude between $|\widetilde{\Psi}_i\rangle_{\text{electron}}$ and $|\widetilde{\Psi}_f\rangle_{\text{electron}}$ is $\langle U_{r'}(\boldsymbol{p}')|S|U_r(\boldsymbol{p})\rangle$, where $S$ is $S$-matrix; however, if the initial and final state of a positron are $|\widetilde{\Psi}_i\rangle_{\text{positron}}=|\widetilde{\Psi}_r(\boldsymbol{p})\rangle_{\text{positron}}$ and $|\widetilde{\Psi}_f\rangle_{\text{positron}}=|\widetilde{\Psi}_{r'}(\boldsymbol{p}')\rangle_{\text{positron}}$, respectively, then according to the Feynman's theory of positrons, the transition amplitude between $|\widetilde{\Psi}_i\rangle_{\text{positron}}$ and $|\widetilde{\Psi}_f\rangle_{\text{positron}}$ is $\langle V_{-r}(-\boldsymbol{p})|S|V_{-r'}(-\boldsymbol{p}')\rangle$, notice that it seems formally as if a negative energy particle transits from the final state $|V_{-r'}(-\boldsymbol{p}')\rangle$ to the initial state $|V_{-r}(-\boldsymbol{p})\rangle$ in inverse time direction.

And, further, if the initial and final state of an electron are

$$|\widetilde{\Psi}_i\rangle_{\text{electron}} = A_1|\widetilde{\Psi}_{r_1}(\boldsymbol{p}_1)\rangle\rangle_{\text{electron}} + A_2|\widetilde{\Psi}_{r_2}(\boldsymbol{p}_2)\rangle\rangle_{\text{electron}},$$

$$|\widetilde{\Psi}_f\rangle_{\text{electron}} = A_1'|\widetilde{\Psi}_{r_1'}(\boldsymbol{p}_1')\rangle\rangle_{\text{electron}} + A_2'|\widetilde{\Psi}_{r_2'}(\boldsymbol{p}_2')\rangle\rangle_{\text{electron}},$$

respectively, where $A_1^* A_1 + A_2^* A_2 = 1$, $A_1'^* A_1' + A_2'^* A_2' = 1$, then the transition amplitude between $|\widetilde{\Psi}_i\rangle_{\text{electron}}$ and $|\widetilde{\Psi}_f\rangle_{\text{electron}}$ is

$$\left(A_1'^*\langle U_{r_1'}(\boldsymbol{p}_1')| + A_2'^*\langle U_{r_2'}(\boldsymbol{p}_2')|\right) S\left(A_1|U_{r_1}(\boldsymbol{p}_1)\rangle + A_2|U_{r_2}(\boldsymbol{p}_2)\rangle\right); \tag{17}$$

however, if the initial and final state of a positron are

$$|\widetilde{\Psi}_i\rangle_{\text{positron}} = B_1|\widetilde{\Psi}_{r_1}(\boldsymbol{p}_1)\rangle\rangle_{\text{positron}} + B_2|\widetilde{\Psi}_{r_2}(\boldsymbol{p}_2)\rangle\rangle_{\text{positron}},$$

$$|\widetilde{\Psi}_f\rangle_{\text{positron}} = B_1'|\widetilde{\Psi}_{r_1'}(\boldsymbol{p}_1')\rangle\rangle_{\text{positron}} + B_2'|\widetilde{\Psi}_{r_2'}(\boldsymbol{p}_2')\rangle\rangle_{\text{positron}},$$

respectively, where $B_1^* B_1 + B_2^* B_2 = 1$, $B_1'^* B_1' + B_2'^* B_2' = 1$, then the transition amplitude between $|\widetilde{\Psi}_i\rangle_{\text{positron}}$ and $|\widetilde{\Psi}_f\rangle_{\text{positron}}$ is

$$\left(B_1^*\langle V_{-r_1}(-\boldsymbol{p}_1)| + B_2^*\langle V_{-r_2}(-\boldsymbol{p}_2)|\right) S\left(B_1'|V_{-r_1'}(-\boldsymbol{p}_1')\rangle + B_2'|V_{-r_2'}(-\boldsymbol{p}_2')\rangle\right). \tag{18}$$

We see that, it seems formally as if a negative energy particle transits from the final state $B_1'|V_{-r_1'}(-\boldsymbol{p}_1')\rangle + B_2'|V_{-r_2'}(-\boldsymbol{p}_2')\rangle$ to the initial state $B_1|V_{-r_1}(-\boldsymbol{p}_1)\rangle + B_2|V_{-r_2}(-\boldsymbol{p}_2)\rangle$ in inverse time direction, hence, the forms appeared in (18) of the coefficients $B_1$ and $B_2$ in the initial state $|\widetilde{\Psi}_i\rangle_{\text{positron}}$ of a positron are $B_1^*$ and $B_2^*$, respectively; this is differ from the case of the coefficients $A_1$ and $A_2$ in the initial state $|\widetilde{\Psi}_i\rangle_{\text{electron}}$ of an electron in (17).

**2.4 Derivation of the calculation rule of probability given by (3)**

According to the calculation rule of probability of many particles system, if a physical quantity $A$ of which the eigenvalue equation is given by (2) is measured for a state of many particles system described by a wave function $\Psi(t; \boldsymbol{x}_1, \boldsymbol{x}_2, \cdots, \boldsymbol{x}_i, \cdots)$, then the probability $P_{\text{system}}(a_n)$ that an eigen-value $a_n$ is obtained is:

$$\begin{aligned} P_{\text{system}}(a_n) = N\int \mathrm{d}^3x_2\mathrm{d}^3x_3\cdots\mathrm{d}^3x_i\cdots \Big[\int \Psi^+(t; \boldsymbol{x}_1', \boldsymbol{x}_2, \cdots, \boldsymbol{x}_i, \cdots)\mathrm{d}^3x_1' A_n(\boldsymbol{x}_1') \\ \times\int A_n^+(\boldsymbol{x}_1)\mathrm{d}^3x_1\Psi(t; \boldsymbol{x}_1, \boldsymbol{x}_2, \cdots, \boldsymbol{x}_i, \cdots)\Big]. \end{aligned} \tag{19}$$

However, if we substitute the wave function (14) to (19), then what result we obtain from (19) includes the observation results about Dirac sea, which is now as the vacuum. On the other hand, *the vacuum is not observable*; thus, an actual probability $P(a_n)$ should be the result after removing the contribution from the vacuum:

$$P(a_n) = P_{\text{system}}(a_n) - P_{\text{vacuum}}(a_n). \tag{20}$$

In (20), $P_{\text{vacuum}}(a_n)$ is the probability that the eigen-value $a_n$ is obtained when the physical quantity $A$ is measured for the vacuum described by the wave function $\Psi_{\text{vacuum}}(\boldsymbol{x}_1, \boldsymbol{x}_2, \cdots, \boldsymbol{x}_i, \cdots)$ given by (11):

$$P_{\text{vacuum}}(a_n) = N\int d^3x_2 d^3x_3 \cdots d^3x_i \cdots \left[\int \Psi^+_{\text{vacuum}}(\boldsymbol{x}'_1, \boldsymbol{x}_2, \cdots, \boldsymbol{x}_i, \cdots) d^3x'_1 A_n(\boldsymbol{x}'_1) \times \int A^+_n(\boldsymbol{x}_1) d^3x_1 \Psi_{\text{vacuum}}(\boldsymbol{x}_1, \boldsymbol{x}_2, \cdots, \boldsymbol{x}_i, \cdots)\right]. \tag{21}$$

The process removing the contribution from the vacuum in (19), and, thus, obtaining (20), is in accord to the spirit of the Dirac hold theory. According to the Dirac hold theory, the total energy and the total charges of the vacuum are $-2\sum_{\boldsymbol{p}}\sqrt{p^2+m^2}$ and $-2Ne$, respectively; if Dirac sea lacks one negative energy state particle with the momentum value $\boldsymbol{p}_0$, then the total energy and the total charges of the vacuum change to $-2\sum_{\boldsymbol{p}}\sqrt{p^2+m^2} - \left(-\sqrt{p_0^2+m^2}\right)$ and $-2Ne-e$, respectively. However, *what we obverse actually are the results after removing the contribution from the vacuum*, hence, what energy and charge we obverse actually are $-2\sum_{\boldsymbol{p}}\sqrt{p^2+m^2} - \left(-\sqrt{p_0^2+m^2}\right) - \left(-2\sum_{\boldsymbol{p}}\sqrt{p^2+m^2}\right) = \sqrt{p_0^2+m^2}$ and $-2Ne-e-(-2Ne) = -e$, respectively; namely, what we obverse actually is an antiparticle with energy $\sqrt{p_0^2+m^2}$ and charge $-e$.

For the wave function (14), using (4) ~ (6), we can prove that (20) is just (3), but the calculation process is lengthy; in occupation number representation, this calculation process can be greatly predigested.

## 3 Discussion in occupation number representation

### 3.1 Some conclusions of many particles system[4]

Considering a set of complete states of orthogonality and normalization $\{|\varphi_i\rangle\}$ and introducing the creation and annihilation operator $\hat{c}_i^+$ and $\hat{c}_i$ of the state $|\varphi_i\rangle$, we have

$$\{\hat{c}_i, \hat{c}_j^+\} = \delta_{ij}, \{\hat{c}_i, \hat{c}_j\} = \{\hat{c}_i^+, \hat{c}_j^+\} = 0, \ \hat{c}_i|0\rangle = 0.$$

Defining

$$\hat{\psi}(\boldsymbol{x}) = \sum_i \langle \boldsymbol{x}|\varphi_i\rangle \hat{c}_i = \sum_i \varphi_i(\boldsymbol{x})\,\hat{c}_i, \ \hat{\psi}^+(\boldsymbol{x}) = \sum_i \langle \varphi_i|\boldsymbol{x}\rangle \hat{c}_i^+ = \sum_i \varphi_i^*(\boldsymbol{x})\,\hat{c}_i^+,$$

$$|\boldsymbol{x}_1, \boldsymbol{x}_2, \cdots, \boldsymbol{x}_N\rangle = \frac{1}{\sqrt{N!}}\hat{\psi}^+(\boldsymbol{x}_1)\hat{\psi}^+(\boldsymbol{x}_2)\cdots\hat{\psi}^+(\boldsymbol{x}_N)|0\rangle,$$

The state vector that $N$ particles are situated $N$ states $|\varphi_{j_1}\rangle, |\varphi_{j_2}\rangle, \cdots, |\varphi_{j_N}\rangle$ is

$$|\Psi(t)\rangle = c_{j_1}^+ c_{j_2}^+ \cdots c_{j_N}^+ |0\rangle,$$

the corresponding wave function $\Psi(t; \boldsymbol{x}_1, \boldsymbol{x}_2, \cdots, \boldsymbol{x}_N)$ is:

$$\begin{aligned}\Psi(t;\boldsymbol{x}_1,\boldsymbol{x}_2,\cdots,\boldsymbol{x}_N) &= \langle \boldsymbol{x}_1,\boldsymbol{x}_2,\cdots,\boldsymbol{x}_N|\Psi(t)\rangle \\ &= \frac{1}{\sqrt{N!}}\sum_{(\boldsymbol{x}_{i_1},\boldsymbol{x}_{i_2},\cdots,\boldsymbol{x}_{i_N})}(-1)^{\delta(\boldsymbol{x}_{i_1},\boldsymbol{x}_{i_2},\cdots,\boldsymbol{x}_{i_N})}\varphi_{j_1}(\boldsymbol{x}_{i_1})\varphi_{j_2}(\boldsymbol{x}_{i_2})\cdots\varphi_{j_N}(\boldsymbol{x}_{i_N}).\end{aligned} \tag{22}$$

The probability $P(\boldsymbol{x}_1,\boldsymbol{x}_2,\cdots,\boldsymbol{x}_N)$ that there is one situated $\boldsymbol{x}_1$, one situated $\boldsymbol{x}_2$, ⋯, one situated $\boldsymbol{x}_N$ in $N$ particles is

$$\begin{aligned}P(\boldsymbol{x}_1,\boldsymbol{x}_2,\cdots,\boldsymbol{x}_N) &= \Psi^*(t;\boldsymbol{x}_1,\boldsymbol{x}_2,\cdots,\boldsymbol{x}_N)\Psi(t;\boldsymbol{x}_1,\boldsymbol{x}_2,\cdots,\boldsymbol{x}_N) \\ &= \langle\Psi(t)|\boldsymbol{x}_1,\boldsymbol{x}_2,\cdots,\boldsymbol{x}_N\rangle\langle\boldsymbol{x}_1,\boldsymbol{x}_2,\cdots,\boldsymbol{x}_N|\Psi(t)\rangle,\end{aligned}$$

and, further, the probability $P(\boldsymbol{x})$ that there is one situated $\boldsymbol{x}$ in the $N$ particles is

$$\begin{aligned}P(\boldsymbol{x}) &= N\int \mathrm{d}^3x_2\mathrm{d}^3x_3\cdots\mathrm{d}^3x_N\Psi^*(t;\boldsymbol{x},\boldsymbol{x}_2,\cdots,\boldsymbol{x}_N)\Psi(t;\boldsymbol{x},\boldsymbol{x}_2,\cdots,\boldsymbol{x}_N) \\ &= N\int \mathrm{d}^3x_2\mathrm{d}^3x_3\cdots\mathrm{d}^3x_N\langle\Psi(t)|\boldsymbol{x},\boldsymbol{x}_2,\cdots,\boldsymbol{x}_N\rangle\langle\boldsymbol{x},\boldsymbol{x}_2,\cdots,\boldsymbol{x}_N|\Psi(t)\rangle \\ &= \langle\Psi(t)|\hat{\psi}^+(\boldsymbol{x})\hat{\psi}(\boldsymbol{x})|\Psi(t)\rangle.\end{aligned}$$

For a physical quantity $A$ of which the eigenvalue equation is given by (2), we define

$$\hat{\psi}(A_n) = \sum_i \langle A_n|\varphi_i\rangle c_i,\ \hat{\psi}^+(A_n) = \sum_i \langle\varphi_i|A_n\rangle c_i^+,$$

$$|A_{n_1},A_{n_2},\cdots,A_{n_N}\rangle = \frac{1}{\sqrt{N!}}\hat{\psi}^+(A_{n_1})\hat{\psi}^+(A_{n_2})\cdots\hat{\psi}^+(A_{n_N})|0\rangle,$$

$$\Psi(t;A_{n_1},A_{n_2},\cdots,A_{n_N}) = \langle A_{n_1},A_{n_2},\cdots,A_{n_N}|\Psi(t)\rangle;$$

The probability $P(A_{n_1},A_{n_2},\cdots,A_{n_N})$ that there is one situated the state $|A_{n_1}\rangle$, one situated $|A_{n_2}\rangle$, ⋯, one situated $|A_{n_N}\rangle$ in $N$ particles is

$$\begin{aligned}P(A_{n_1},A_{n_2},\cdots,A_{n_N}) &= \Psi^*(t;A_{n_1},A_{n_2},\cdots,A_{n_N})\Psi(t;A_{n_1},A_{n_2},\cdots,A_{n_N}) \\ &= \langle\Psi(t)|A_{n_1},A_{n_2},\cdots,A_{n_N}\rangle\langle A_{n_1},A_{n_2},\cdots,A_{n_N}|\Psi(t)\rangle,\end{aligned} \tag{23}$$

and, further, the probability $P(A_n)$ that there is one situated the state $|A_n\rangle$ in the $N$ particles is

$$\begin{aligned}P(A_n) &= N\sum_{n_2}\cdots\sum_{n_N}\Psi^*(t;A_n,A_{n_2},\cdots,A_{n_N})\Psi(t;A_n,A_{n_2},\cdots,A_{n_N}) \\ &= N\sum_{n_2}\cdots\sum_{n_N}\langle\Psi(t)|A_n,A_{n_2},\cdots,A_{n_N}\rangle\langle A_n,A_{n_2},\cdots,A_{n_N}|\Psi(t)\rangle \\ &= \langle\Psi(t)|\hat{\psi}^+(A_n)\hat{\psi}(A_n)|\Psi(t)\rangle.\end{aligned} \tag{24}$$

**3.2 The case of the Dirac theory**

Introducing the creation and annihilation operator $\hat{a}_r^+(\boldsymbol{p})$ and $\hat{a}_r(\boldsymbol{p})$ of the state $|U_r(\boldsymbol{p})\rangle$, and the creation and annihilation operator $\hat{b}_r^+(\boldsymbol{p})$ and $\hat{b}_r(\boldsymbol{p})$ of the state $|V_r(\boldsymbol{p})\rangle$, where the states $\left\{|U_r(\boldsymbol{p})\rangle, |V_r(\boldsymbol{p})\rangle, r = -\frac{1}{2},\frac{1}{2}\right\}$ satisfy (4) ~ (6), the anticommutation relations among $\hat{a}_r(\boldsymbol{p})$, $\hat{a}_r^+(\boldsymbol{p})$, $\hat{b}_r(\boldsymbol{p})$ and $\hat{b}_r^+(\boldsymbol{p})$ are

$$\left\{\hat{a}_r(\boldsymbol{p}), \hat{a}_{r'}^+(\boldsymbol{p}')\right\} = \delta_{rr'}\delta^3(\boldsymbol{p}-\boldsymbol{p}'),\ \left\{\hat{b}_r(\boldsymbol{p}), \hat{b}_{r'}^+(\boldsymbol{p}')\right\} = \delta_{rr'}\delta^3(\boldsymbol{p}-\boldsymbol{p}'), \quad \text{Others equal zero.} \tag{25}$$

According to the idea of Dirac sea, the state of the vacuum can be described by

$$|0\rangle = b^+_{-\frac{1}{2}}(\boldsymbol{p}_1) b^+_{\frac{1}{2}}(\boldsymbol{p}_1) b^+_{-\frac{1}{2}}(\boldsymbol{p}_2) b^+_{\frac{1}{2}}(\boldsymbol{p}_2) \cdots b^+_{-\frac{1}{2}}(\boldsymbol{p}_i) b^+_{\frac{1}{2}}(\boldsymbol{p}_i) \cdots |\text{bare vacuum}\rangle , \tag{26}$$

where $|\text{bare vacuum}\rangle$ is the state vector of "the bare vacuum" in which there is not any particle possessing positive or negative energy.

Defining

$$\hat{d}_r(\boldsymbol{p}) = \hat{b}_r^+(\boldsymbol{p}),\ \hat{d}_r^+(\boldsymbol{p}) = \hat{b}_r(\boldsymbol{p}), \tag{27}$$

we can obtain the anticommutation relations among $\hat{a}_r(\boldsymbol{p})$, $\hat{a}_r^+(\boldsymbol{p})$, $\hat{d}_r(\boldsymbol{p})$ and $\hat{d}_r^+(\boldsymbol{p})$ from (25):

$$\left\{\hat{a}_r(\boldsymbol{p}), \hat{a}_{r'}^+(\boldsymbol{p}')\right\} = \delta_{rr'}\delta^3(\boldsymbol{p}-\boldsymbol{p}'),\ \left\{\hat{d}_r(\boldsymbol{p}), \hat{d}_{r'}^+(\boldsymbol{p}')\right\} = \delta_{rr'}\delta^3(\boldsymbol{p}-\boldsymbol{p}'), \quad \text{Others equal zero.} \tag{28}$$

and we have

$$\hat{a}_r(\boldsymbol{p})|0\rangle = 0,\ \hat{d}_r(\boldsymbol{p})|0\rangle = 0 . \tag{29}$$

According to some conclusions given by §3.1, we define

$$\begin{aligned} \hat{\psi}(\boldsymbol{x}) &= \int \mathrm{d}^3 p \sum_{r=1}^{2} \left( \langle \boldsymbol{x} | U_r(\boldsymbol{p}) \rangle \hat{a}_r(\boldsymbol{p}) + \langle \boldsymbol{x} | V_r(\boldsymbol{p}) \rangle \hat{b}_r(\boldsymbol{p}) \right) \\ &= \int \mathrm{d}^3 p \sum_{r=1}^{2} \left( U_r(\boldsymbol{p},\boldsymbol{x}) \hat{a}_r(\boldsymbol{p}) + V_r(\boldsymbol{p},\boldsymbol{x}) \hat{d}_r^+(\boldsymbol{p}) \right), \end{aligned} \tag{30}$$

$$\begin{aligned} \hat{\psi}^+(\boldsymbol{x}) &= \int \mathrm{d}^3 p \sum_{r=1}^{2} \left( \langle U_r(\boldsymbol{p}) | \boldsymbol{x} \rangle \hat{a}_r^+(\boldsymbol{p}) + \langle V_r(\boldsymbol{p}) | \boldsymbol{x} \rangle \hat{b}_r^+(\boldsymbol{p}) \right) \\ &= \int \mathrm{d}^3 p \sum_{r=1}^{2} \left( U_r^*(\boldsymbol{p},\boldsymbol{x}) \hat{a}_r^+(\boldsymbol{p}) + U_r^*(\boldsymbol{p},\boldsymbol{x}) \hat{d}_r(\boldsymbol{p}) \right), \end{aligned} \tag{31}$$

$$|\boldsymbol{x}_1, \boldsymbol{x}_2, \cdots, \boldsymbol{x}_i, \cdots\rangle = \frac{1}{\sqrt{N!}} \hat{\psi}^+(\boldsymbol{x}_1) \hat{\psi}^+(\boldsymbol{x}_2) \cdots \hat{\psi}^+(\boldsymbol{x}_i) \cdots |\text{ bare vacuum}\rangle ; \tag{32}$$

Based on above $\hat{\psi}(\boldsymbol{x})$, $\hat{\psi}^+(\boldsymbol{x})$ and $|\boldsymbol{x}_1, \boldsymbol{x}_2, \cdots, \boldsymbol{x}_i, \cdots\rangle$, according to (22) in §3.1, (12), (13) and (14) can be written to the form

$$\Psi_r^{(+)}(\boldsymbol{p}; \boldsymbol{p}_1, \boldsymbol{p}_2, \cdots, \boldsymbol{p}_{i'}, \cdots; \boldsymbol{x}_1, \boldsymbol{x}_2, \cdots, \boldsymbol{x}_i, \cdots) = \langle \boldsymbol{x}_1, \boldsymbol{x}_2, \cdots, \boldsymbol{x}_i, \cdots | \Psi_r^{(+)}(\boldsymbol{p}) \rangle , \tag{33}$$

$$\Psi_r^{(-)}(\boldsymbol{p}; \boldsymbol{p}_1, \boldsymbol{p}_2, \cdots, \boldsymbol{p}_{i'}, \cdots; \boldsymbol{x}_1, \boldsymbol{x}_2, \cdots, \boldsymbol{x}_i, \cdots) = \langle \boldsymbol{x}_1, \boldsymbol{x}_2, \cdots, \boldsymbol{x}_i, \cdots | \Psi_r^{(-)}(\boldsymbol{p}) \rangle , \tag{34}$$

$$\Psi(t; \boldsymbol{x}_1, \boldsymbol{x}_2, \cdots, \boldsymbol{x}_i, \cdots) = \langle \boldsymbol{x}_1, \boldsymbol{x}_2, \cdots, \boldsymbol{x}_i, \cdots | \Psi(t) \rangle . \tag{35}$$

where

$$|\Psi_r^{(+)}(\boldsymbol{p})\rangle = \hat{a}_r^+(\boldsymbol{p})|0\rangle\,, \quad |\Psi_r^{(-)}(\boldsymbol{p})\rangle = \hat{b}_r(\boldsymbol{p})|0\rangle = \hat{d}_r^+(\boldsymbol{p})|0\rangle\,, \tag{36}$$

$$\begin{aligned} |\Psi(t)\rangle &= \int \mathrm{d}^3 p \sum_{r=1}^{2} \Big( A_r(t\,,\boldsymbol{p})\hat{a}_r^+(\boldsymbol{p})|0\rangle + B_r^*(t\,,\boldsymbol{p})\hat{b}_r(\boldsymbol{p})|0\rangle \Big) \\ &= \int \mathrm{d}^3 p \sum_{r=1}^{2} \Big( \langle U_r(\boldsymbol{p})|\widetilde{\Psi}(t)\rangle \hat{a}_r^+(\boldsymbol{p})|0\rangle + \langle \widetilde{\Psi}(t)|V_r(\boldsymbol{p})\rangle \hat{d}_r^+(\boldsymbol{p})|0\rangle \Big). \end{aligned} \tag{37}$$

In (37), we have used (9). Notice that the state vector $|\Psi(t)\rangle$ of many particles system is differ from the state vector $|\widetilde{\Psi}(t)\rangle$ of "one-particle" of which the correcponding wave function is given by (8).

For a physical quantity $A$ of which the eigenvalue equation is given by (2), we define

$$\begin{aligned} \hat{\psi}(A_n) &= \int \mathrm{d}^3 p \sum_{r=1}^{2} \Big( \langle A_n|U_r(\boldsymbol{p})\rangle \hat{a}_r(\boldsymbol{p}) + \langle A_n|V_r(\boldsymbol{p})\rangle \hat{b}_r(\boldsymbol{p}) \Big) \\ &= \int \mathrm{d}^3 p \sum_{r=1}^{2} \Big( \langle A_n|U_r(\boldsymbol{p})\rangle \hat{a}_r(\boldsymbol{p}) + \langle A_n|V_r(\boldsymbol{p})\rangle \hat{d}_r^+(\boldsymbol{p}) \Big), \end{aligned} \tag{38}$$

$$\begin{aligned} \hat{\psi}^+(A_n) &= \int \mathrm{d}^3 p \sum_{r=1}^{2} \Big( \langle U_r(\boldsymbol{p})|A_n\rangle \hat{a}_r^+(\boldsymbol{p}) + \langle V_r(\boldsymbol{p})|A_n\rangle \hat{b}_r^+(\boldsymbol{p}) \Big) \\ &= \int \mathrm{d}^3 p \sum_{r=1}^{2} \Big( \langle U_r(\boldsymbol{p})|A_n\rangle \hat{a}_r^+(\boldsymbol{p}) + \langle V_r(\boldsymbol{p})|A_n\rangle \hat{d}_r(\boldsymbol{p}) \Big), \end{aligned} \tag{39}$$

$$|A_{n_1}, A_{n_2}, \cdots, A_{n_i}, \cdots\rangle = \frac{1}{\sqrt{N!}} \hat{\psi}^+(A_{n_1}) \hat{\psi}^+(A_{n_2}) \cdots \hat{\psi}^+(A_{n_i}) \cdots |\,\text{bare vacuum}\rangle\,. \tag{40}$$

The form of (23) in §3.1 reads now

$$\begin{aligned} P_{\text{system}}(A_n) &= \Psi^*(t\,; A_n, A_{n_2}, \cdots, A_{n_i}, \cdots) \Psi(t\,; A_n, A_{n_2}, \cdots, A_{n_i}, \cdots) \\ &= \langle \Psi(t)|A_n, A_{n_2}, \cdots, A_{n_i}, \cdots\rangle\langle A_n, A_{n_2}, \cdots, A_{n_i}, \cdots|\Psi(t)\rangle\,. \end{aligned} \tag{41}$$

According to (24) in §3.1, (19) and (21) now can be written to the forms

$$\begin{aligned} P_{\text{system}}(A_n) &= N \sum_{n_2} \cdots \sum_{n_N} \Psi^*(t\,; A_n, A_{n_2}, \cdots, A_{n_i}, \cdots) \Psi(t\,; A_n, A_{n_2}, \cdots, A_{n_i}, \cdots) \\ &= N \sum_{n_2} \cdots \sum_{n_N} \langle \Psi(t)|A_n, A_{n_2}, \cdots, A_{n_i}, \cdots\rangle\langle A_n, A_{n_2}, \cdots, A_{n_i}, \cdots|\Psi(t)\rangle \\ &= \langle \Psi(t)|\hat{\psi}^+(A_n)\, \hat{\psi}(A_n)\, |\Psi(t)\rangle\,. \end{aligned} \tag{42}$$

$$\begin{aligned} P_{\text{vacuum}}(A_n) &= N \sum_{n_2} \cdots \sum_{n_N} \Psi^*_{\text{vacuum}}(t\,; A_n, A_{n_2}, \cdots, A_{n_i}, \cdots) \Psi_{\text{vacuum}}(t\,; A_n, A_{n_2}, \cdots, A_{n_i}, \cdots) \\ &= N \sum_{n_2} \cdots \sum_{n_N} \langle 0|A_n, A_{n_2}, \cdots, A_{n_i}, \cdots\rangle\langle A_n, A_{n_2}, \cdots, A_{n_i}, \cdots|0\rangle \\ &= \langle 0|\hat{\psi}^+(A_n)\, \hat{\psi}(A_n)\, |0\rangle\,. \end{aligned} \tag{43}$$

Substituting (37), (30) and (31) to (42) and (43), using the characteristics given by (6), (9), (10), (28) and (29), we obtain

$$
\begin{aligned}
P_{\text{system}}(A_n) &= \int \mathrm{d}^3 p \sum_{r=1}^{2} \left(A_r^*(t\,,\boldsymbol{p})\langle U_r(\boldsymbol{p})|A_n\rangle\right)\cdot \int \mathrm{d}^3 p' \sum_{r'=1}^{2} \left(\langle A_n|U_{r'}(\boldsymbol{p}')\rangle A_{r'}(t\,,\boldsymbol{p}')\right) \\
&\quad - \int \mathrm{d}^3 p \sum_{r=1}^{2} \left(B_r(t\,,\boldsymbol{p})\langle A_n|V_r(\boldsymbol{p})\rangle\right)\cdot \int \mathrm{d}^3 p' \sum_{r'=1}^{2} \left(\langle V_{r'}(\boldsymbol{p}')|A_n\rangle B_{r'}^*(t\,,\boldsymbol{p}')\right) \\
&\quad + \int \mathrm{d}^3 p \sum_{r=1}^{2} \left(A_r^*(t\,,\boldsymbol{p})A_r(t\,,\boldsymbol{p}) + B_r(t\,,\boldsymbol{p})B_r^*(t\,,\boldsymbol{p})\right)\cdot \int \mathrm{d}^3 p' \sum_{r'=1}^{2} \left(\langle A_n|V_{r'}(\boldsymbol{p}')\rangle\langle V_{r'}(\boldsymbol{p}')|A_n\rangle\right) \\
&= \int \mathrm{d}^3 p \sum_{r=1}^{2} \left(\langle\widetilde{\Psi}(t)|U_r(\boldsymbol{p})\rangle\langle U_r(\boldsymbol{p})|A_n\rangle\right)\cdot \int \mathrm{d}^3 p' \sum_{r'=1}^{2} \left(\langle A_n|U_{r'}(\boldsymbol{p}')\rangle\langle U_{r'}(\boldsymbol{p}')|\widetilde{\Psi}(t)\rangle\right) \qquad (44) \\
&\quad - \int \mathrm{d}^3 p \sum_{r=1}^{2} \left(\langle A_n|V_r(\boldsymbol{p})\rangle\langle V_r(\boldsymbol{p})|\widetilde{\Psi}(t)\rangle\right)\cdot \int \mathrm{d}^3 p' \sum_{r'=1}^{2} \left(\langle\widetilde{\Psi}(t)|V_{r'}(\boldsymbol{p}')\rangle\langle V_{r'}(\boldsymbol{p}')|A_n\rangle\right) \\
&\quad + \int \mathrm{d}^3 p \sum_{r=1}^{2} \left(\langle A_n|V_r(\boldsymbol{p})\rangle\langle V_r(\boldsymbol{p})|A_n\rangle\right) \\
&= \langle\widetilde{\Psi}|\hat{M}_+|A_n\rangle\langle A_n|\hat{M}_+|\widetilde{\Psi}\rangle - \langle\widetilde{\Psi}|\hat{M}_-|A_n\rangle\langle A_n|\hat{M}_-|\widetilde{\Psi}\rangle + \langle A_n|\hat{M}_-|A_n\rangle\,,
\end{aligned}
$$

$$
P_{\text{vacuum}}(A_n) = \langle 0|\hat{\psi}^+(A_n)\,\hat{\psi}(A_n)\,|0\rangle = \langle A_n|\hat{M}_-|A_n\rangle\;; \qquad (45)
$$

Finally, we obtain

$$
P(A_n) = P_{\text{system}}(A_n) - P_{\text{vacuum}}(A_n) = \langle\Psi|\hat{M}_+|A_n\rangle\langle A_n|\hat{M}_+|\Psi\rangle - \langle\Psi|\hat{M}_-|A_n\rangle\langle A_n|\hat{M}_-|\Psi\rangle\,. \qquad (46)
$$

This is just (3).

### 3.3 Three notes

① It is obvious that the calculation result of the rule (3) may be negative. For example, if we use (3) to calculate the the probability that a negative energy particle lacks at point $x$, e.g., an antiparticle situated $x$, we obtain a "negative probability". Hence, what result obtained by (3) should be as "charge density" but not "particle number density". In fact, for relativistic case, we can define *local charge state* but can not define *local particle number state* [5]. On the other hand, if a particle is situated in an area that is less than the Compton wavelength of the particle, then maybe the corresponding antiparticles are created. Hence, we can only study *local charge density* but can not study *local particle number density* for relativistic case.

② The calculation rule of probability (3) can be generalized to many particles case by using the method given by §3.1. For example, according to Ref. [4], the probability $P(\boldsymbol{x}_1\,,\boldsymbol{x}_2)$ that there is one situated $\boldsymbol{x}_1$ and other situated $\boldsymbol{x}_2$ in the $N$ particles is

$$
\begin{aligned}
P_{\text{system}}(\boldsymbol{x}_1\,,\boldsymbol{x}_2) &= N(N-1)\int \mathrm{d}^3x_3\cdots\mathrm{d}^3x_i\cdots\langle\Psi(t)|\boldsymbol{x}_1\,,\boldsymbol{x}_2\,,\boldsymbol{x}_3\,,\cdots,\boldsymbol{x}_i\,,\cdots\rangle\langle\boldsymbol{x}_1\,,\boldsymbol{x}_2\,,\boldsymbol{x}_3\,,\cdots,\boldsymbol{x}_i\,,\cdots|\Psi(t)\rangle \\
&= \langle\Psi(t)|\hat{\psi}^+(\boldsymbol{x}_2)\hat{\psi}^+(\boldsymbol{x}_1)\hat{\psi}(\boldsymbol{x}_1)\hat{\psi}(\boldsymbol{x}_2)|\Psi(t)\rangle\,,
\end{aligned}
$$

after removing the correcspongding contribution from the vacuum, the actual probability $P(\boldsymbol{x}_1\,,\boldsymbol{x}_2)$ is

$$
P(\boldsymbol{x}_1\,,\boldsymbol{x}_2) = \langle\Psi(t)|\hat{\psi}^+(\boldsymbol{x}_2)\hat{\psi}^+(\boldsymbol{x}_1)\hat{\psi}(\boldsymbol{x}_1)\hat{\psi}(\boldsymbol{x}_2)|\Psi(t)\rangle - \langle 0|\hat{\psi}^+(\boldsymbol{x}_2)\hat{\psi}^+(\boldsymbol{x}_1)\hat{\psi}(\boldsymbol{x}_1)\hat{\psi}(\boldsymbol{x}_2)|0\rangle\,.
$$

③ If we want to calculate the probability observing some particles situated some eigenstates of momentum, e.g., $\hat{a}_r^+(\boldsymbol{p})|0\rangle$, $\hat{a}_r^+(\boldsymbol{p})\hat{d}_r^+(\boldsymbol{p})|0\rangle$, etc, notice that according to (26), $|0\rangle$ is a many particle system, then we must use (41). For example, if we calculate the probability that a system described by state vector $|\Psi(t)\rangle$ is situated $\hat{a}_r^+(\boldsymbol{p})|0\rangle$, then we cannot use (42) but must use (41), the result is $P_{\text{system}} = \langle\Psi(t)|\hat{a}_r^+(\boldsymbol{p})|0\rangle\langle 0|\hat{a}_r(\boldsymbol{p})|\Psi(t)\rangle$; of course, we must remove the correcspongding contribution from the vacuum, however, from (29) we have $P_{\text{vacuum}} = \langle 0|\hat{a}_r^+(\boldsymbol{p})|0\rangle\langle 0|\hat{a}_r(\boldsymbol{p})|0\rangle = 0$. We see that, for the case of calculating the probability observing some particles situated some eigenstates of momentum, what results we obtain are the same as that of the quantum theory of field.

## 4 The case of particle satisfying the Klein-Gordon equation

As is well known, if the wave function $\varphi(t,\boldsymbol{x})$ of a particle satisfies the Klein-Gordon equation:

$$-\frac{\partial^2\varphi(t,\boldsymbol{x})}{\partial^2 t} + \nabla^2\varphi(t,\boldsymbol{x}) - m^2\varphi(t,\boldsymbol{x}) = 0, \qquad (47)$$

then position probability density and three dimentional current density are

$$\rho(t,\boldsymbol{x}) = \frac{\mathrm{i}}{2m}\left(\varphi^*(t,\boldsymbol{x})\frac{\partial\varphi(t,\boldsymbol{x})}{\partial t} - \varphi(t,\boldsymbol{x})\frac{\partial\varphi^*(t,\boldsymbol{x})}{\partial t}\right), \qquad (48)$$

$$\boldsymbol{j}(t,\boldsymbol{x}) = -\frac{\mathrm{i}}{2m}\left(\varphi^*(t,\boldsymbol{x})\nabla\varphi(t,\boldsymbol{x}) - \varphi(t,\boldsymbol{x})\nabla\varphi^*(t,\boldsymbol{x})\right). \qquad (49)$$

We consider the following wave function saticfying the Klein-Gordon equation:

$$\varphi(t,\boldsymbol{x}) = \int\frac{\mathrm{d}^3k}{(2\pi)^3}\frac{m}{\sqrt{k^2+m^2}}\mathrm{e}^{-\mathrm{i}\sqrt{k^2+m^2}(t-t_0)}\mathrm{e}^{\mathrm{i}\boldsymbol{k}\cdot(\boldsymbol{x}-\boldsymbol{x}_0)}, \qquad (50)$$

for which according to (48) we have

$$\begin{aligned}\rho(t,\boldsymbol{x}) = \frac{\mathrm{i}}{2m}\Bigg(&\int\frac{\mathrm{d}^3k}{(2\pi)^3}\frac{m}{\sqrt{k^2+m^2}}\mathrm{e}^{\mathrm{i}\sqrt{k^2+m^2}(t-t_0)}\mathrm{e}^{-\mathrm{i}\boldsymbol{k}\cdot(\boldsymbol{x}-\boldsymbol{x}_0)}\cdot\int\frac{\mathrm{d}^3k'}{(2\pi)^3}(-\mathrm{i}m)\mathrm{e}^{-\mathrm{i}\sqrt{k'^2+m^2}(t-t_0)}\mathrm{e}^{\mathrm{i}\boldsymbol{k}'\cdot(\boldsymbol{x}-\boldsymbol{x}_0)}\\ &-\int\frac{\mathrm{d}^3k}{(2\pi)^3}\frac{m}{\sqrt{k^2+m^2}}\mathrm{e}^{-\mathrm{i}\sqrt{k^2+m^2}(t-t_0)}\mathrm{e}^{\mathrm{i}\boldsymbol{k}\cdot(\boldsymbol{x}-\boldsymbol{x}_0)}\cdot\int\frac{\mathrm{d}^3k'}{(2\pi)^3}(\mathrm{i}m)\mathrm{e}^{\mathrm{i}\sqrt{k'^2+m^2}(t-t_0)}\mathrm{e}^{-\mathrm{i}\boldsymbol{k}'\cdot(\boldsymbol{x}-\boldsymbol{x}_0)}\Bigg),\end{aligned} \qquad (51)$$

and we can prove easily that

$$\rho(t,\boldsymbol{x}) = I_1(t-\boldsymbol{x}, t_0-\boldsymbol{x}_0) = \rho_2(t,\boldsymbol{x};t_0,\boldsymbol{x}_0), \qquad (52)$$

we see that such $\rho(t,\boldsymbol{x})$ is in accord with Tab. 2.

Substituting (50) to (49) we have

$$\begin{aligned}\boldsymbol{j}(t,\boldsymbol{x}) = -\frac{\mathrm{i}}{2m}\Bigg(&\int\frac{\mathrm{d}^3k}{(2\pi)^3}\frac{m}{\sqrt{k^2+m^2}}\mathrm{e}^{\mathrm{i}\sqrt{k^2+m^2}(t-t_0)}\mathrm{e}^{-\mathrm{i}\boldsymbol{k}\cdot(\boldsymbol{x}-\boldsymbol{x}_0)}\nabla\int\frac{\mathrm{d}^3k'}{(2\pi)^3}\frac{m}{\sqrt{k'^2+m^2}}\mathrm{e}^{-\mathrm{i}\sqrt{k'^2+m^2}(t-t_0)}\mathrm{e}^{\mathrm{i}\boldsymbol{k}'\cdot(\boldsymbol{x}-\boldsymbol{x}_0)}\\ &-\int\frac{\mathrm{d}^3k}{(2\pi)^3}\frac{m}{\sqrt{k^2+m^2}}\mathrm{e}^{-\mathrm{i}\sqrt{k^2+m^2}(t-t_0)}\mathrm{e}^{\mathrm{i}\boldsymbol{k}\cdot(\boldsymbol{x}-\boldsymbol{x}_0)}\nabla\int\frac{\mathrm{d}^3k'}{(2\pi)^3}\frac{m}{\sqrt{k'^2+m^2}}\mathrm{e}^{\mathrm{i}\sqrt{k'^2+m^2}(t-t_0)}\mathrm{e}^{-\mathrm{i}\boldsymbol{k}'\cdot(\boldsymbol{x}-\boldsymbol{x}_0)}\Bigg),\end{aligned} \qquad (53)$$

we can prove that (See the Appendix of this paper)

$$\boldsymbol{j}(t,\boldsymbol{x}) = \boldsymbol{j}_2(t,\boldsymbol{x}). \qquad (54)$$

Hence, using (48) and (49), for the wave function (50), we can obtaion the conclusions being in accord with Tab. 2.

On the other hand, at time $t_0$, (50) becomes

$$\varphi(t_0, \boldsymbol{x}) = \int \frac{\mathrm{d}^3 k}{(2\pi)^3} \frac{m}{\sqrt{k^2 + m^2}} \mathrm{e}^{\mathrm{i}\boldsymbol{k}\cdot(\boldsymbol{x}-\boldsymbol{x}_0)}; \tag{55}$$

The form of (55) is different from the wave function $\Psi(\boldsymbol{x}) = \delta^3(\boldsymbol{x} - \boldsymbol{x}_0)$ that describes a particle situated at $\boldsymbol{x}_0$. However, if we use (48) to calculate the probability $\rho(t_0, \boldsymbol{x})$ that a particle described by the wave function (55) is situated $\boldsymbol{x}$ at time $t_0$, according to (51) we have

$$\begin{aligned}
\rho(t_0, \boldsymbol{x}) &= \lim_{t\to t_0} \rho(t, \boldsymbol{x}) \\
&= \frac{\mathrm{i}}{2m}\left( \int \frac{\mathrm{d}^3 k}{(2\pi)^3} \frac{m}{\sqrt{k^2+m^2}} \mathrm{e}^{-\mathrm{i}\boldsymbol{k}\cdot(\boldsymbol{x}-\boldsymbol{x}_0)} \cdot \int \frac{\mathrm{d}^3 k'}{(2\pi)^3} (-\mathrm{i}m) \mathrm{e}^{\mathrm{i}\boldsymbol{k}'\cdot(\boldsymbol{x}-\boldsymbol{x}_0)} \right. \\
&\quad \left. - \int \frac{\mathrm{d}^3 k}{(2\pi)^3} \frac{m}{\sqrt{k^2+m^2}} \mathrm{e}^{\mathrm{i}\boldsymbol{k}\cdot(\boldsymbol{x}-\boldsymbol{x}_0)} \cdot \int \frac{\mathrm{d}^3 k'}{(2\pi)^3} (\mathrm{i}m) \mathrm{e}^{-\mathrm{i}\boldsymbol{k}'\cdot(\boldsymbol{x}-\boldsymbol{x}_0)} \right) \\
&= \int \frac{\mathrm{d}^3 k}{(2\pi)^3} \frac{m}{\sqrt{k^2+m^2}} \delta^3(\boldsymbol{x}-\boldsymbol{x}_0) = \delta^3(0)\delta^3(\boldsymbol{x}-\boldsymbol{x}_0).
\end{aligned} \tag{56}$$

In the last step of the above calculation process, we have used

$$\delta^3(0) = \lim_{\boldsymbol{x}\to\boldsymbol{x}_0} \delta^3(\boldsymbol{x}-\boldsymbol{x}_0) = \lim_{\boldsymbol{x}\to\boldsymbol{x}_0} \int \frac{\mathrm{d}^3 k'}{(2\pi)^3} \mathrm{e}^{\mathrm{i}\boldsymbol{k}'\cdot(\boldsymbol{x}-\boldsymbol{x}_0)} = \int \frac{\mathrm{d}^3 k'}{(2\pi)^3} = \int \frac{\mathrm{d}^3 k}{(2\pi)^3} \frac{m}{\sqrt{k^2+m^2}},$$

where $\boldsymbol{k}' = K(k)\boldsymbol{k}$, $K(k)$ satisfies $\dfrac{\mathrm{d}\left\{[K(k)k]^3\right\}}{\mathrm{d}k} = \dfrac{3mk^2}{\sqrt{k^2+m^2}}$.

The result of (56) is in accord with the form of "position probability density". Because, according to the normal calculation rule of probability $P(a_n) = \langle\Psi|A_n\rangle\langle A_n|\Psi\rangle$, the "position probability density" that a particle described by the wave function $\Psi(\boldsymbol{x}) = \delta^3(\boldsymbol{x}-\boldsymbol{x}_0)$ and situated $\boldsymbol{x}$ is $\Psi^+(\boldsymbol{x})\Psi(\boldsymbol{x}) = \delta^3(\boldsymbol{x}-\boldsymbol{x}_0)\delta^3(\boldsymbol{x}-\boldsymbol{x}_0) = \delta^3(0)\delta^3(\boldsymbol{x}-\boldsymbol{x}_0)$. Hence, under the calculation rule of position probability density (48), the result (56) shows that the wave function (55) can describe the state of particle satisfying the Klein-Gordon equation and situated $\boldsymbol{x}_0$ at time $t_0$.

We can generalize the formula (48) of calculating position probability density to the case of arbitrary physical quantity $A$ and obtain a calculation rule of the probability that an eigen-value $a_n$ is observed when the physical quantity $A$ of which the eigenvalue equation is given by (2) is measured for a state described by the state vector $|\varphi(t)\rangle$ satisfying the Klein-Gordon equation:

$$\rho(t, a_n) = \frac{\mathrm{i}}{2m}\left( \langle\varphi(t)|A_n\rangle \frac{\partial\langle A_n|\varphi(t)\rangle}{\partial t} - \frac{\partial\langle\varphi(t)|A_n\rangle}{\partial t}\langle A_n|\varphi(t)\rangle \right), \tag{57}$$

where

$$\langle A_n|\varphi(t)\rangle = \int \mathrm{d}^3 x \langle A_n|\boldsymbol{x}\rangle\langle\boldsymbol{x}|\varphi(t)\rangle = \int \mathrm{d}^3 x A_n^*(\boldsymbol{x})\varphi(t, \boldsymbol{x}). \tag{58}$$

Especially, for momentum, (57) and (58) become

$$\rho(t,\boldsymbol{p})=\frac{\mathrm{i}}{2m}\left(\varphi^*(t,\boldsymbol{p})\frac{\partial\varphi(t,\boldsymbol{p})}{\partial t}-\varphi(t,\boldsymbol{p})\frac{\partial\varphi^*(t,\boldsymbol{p})}{\partial t}\right), \tag{59}$$

$$\varphi(t,\boldsymbol{p})=\int \mathrm{d}^3x\frac{1}{(2\pi)^{3/2}}\mathrm{e}^{-\mathrm{i}\boldsymbol{p}\cdot\boldsymbol{x}}\varphi(t,\boldsymbol{x}). \tag{60}$$

For the wave function (50), substituting it to (60) we have

$$\varphi(t,\boldsymbol{p})=\frac{1}{(2\pi)^{3/2}}\frac{m}{\sqrt{p^2+m^2}}\mathrm{e}^{-\mathrm{i}\sqrt{p^2+m^2}(t-t_0)}\mathrm{e}^{-\mathrm{i}\boldsymbol{p}\cdot\boldsymbol{x}_0}\,; \tag{61}$$

and, further, substituting (61) to (59), we obtain

$$\rho(t,\boldsymbol{p})=\frac{1}{(2\pi)^3}\frac{m}{\sqrt{p^2+m^2}}=f_2(\boldsymbol{p}). \tag{61}$$

According to the above discussion and the conclusions (61), (52) and (54), we see that if we use the calculation rule of probability (57) for particle satisfying the Klein-Gordon equation, then we can obtain conclusions being in accord with Tab. 2.

Of course, as is well known, what we obtain from (48) should be as “local charge density”.

## Appendix

We use the method given in Ref. [1] that prove the three mathematical formulas in §1.3 to prove (50). For the sake of brevity, we rewrite (53) to the from

$$\boldsymbol{j}(t,\boldsymbol{x})=\boldsymbol{J}(t-t_0,\boldsymbol{x}-\boldsymbol{x}_0), \tag{A-1}$$

where

$$\begin{aligned}\boldsymbol{J}(t,\boldsymbol{x})=&-\frac{\mathrm{i}}{2m}\left(\int\frac{\mathrm{d}^3k}{(2\pi)^3}\frac{m}{\sqrt{k^2+m^2}}\mathrm{e}^{\mathrm{i}\sqrt{k^2+m^2}t}\mathrm{e}^{-\mathrm{i}\boldsymbol{k}\cdot\boldsymbol{x}}\,\nabla\int\frac{\mathrm{d}^3k'}{(2\pi)^3}\frac{m}{\sqrt{k'^2+m^2}}\mathrm{e}^{-\mathrm{i}\sqrt{k'^2+m^2}t}\mathrm{e}^{\mathrm{i}\boldsymbol{k}'\cdot\boldsymbol{x}}\right.\\&\left.-\int\frac{\mathrm{d}^3k}{(2\pi)^3}\frac{m}{\sqrt{k^2+m^2}}\mathrm{e}^{-\mathrm{i}\sqrt{k^2+m^2}t}\mathrm{e}^{\mathrm{i}\boldsymbol{k}\cdot\boldsymbol{x}}\,\nabla\int\frac{\mathrm{d}^3k'}{(2\pi)^3}\frac{m}{\sqrt{k'^2+m^2}}\mathrm{e}^{\mathrm{i}\sqrt{k'^2+m^2}t}\mathrm{e}^{-\mathrm{i}\boldsymbol{k}'\cdot\boldsymbol{x}}\right)\\=&-\frac{1}{m}\left(\int\frac{\mathrm{d}^3k}{(2\pi)^3}\frac{m}{\sqrt{k^2+m^2}}\cos\left(\sqrt{k^2+m^2}t\right)\mathrm{e}^{-\mathrm{i}\boldsymbol{k}\cdot\boldsymbol{x}}\,\nabla\int\frac{\mathrm{d}^3k'}{(2\pi)^3}\frac{m}{\sqrt{k'^2+m^2}}\sin\left(\sqrt{k'^2+m^2}t\right)\mathrm{e}^{\mathrm{i}\boldsymbol{k}'\cdot\boldsymbol{x}}\right.\\&\left.-\int\frac{\mathrm{d}^3k}{(2\pi)^3}\frac{m}{\sqrt{k^2+m^2}}\sin\left(\sqrt{k^2+m^2}t\right)\mathrm{e}^{\mathrm{i}\boldsymbol{k}\cdot\boldsymbol{x}}\,\nabla\int\frac{\mathrm{d}^3k'}{(2\pi)^3}\frac{m}{\sqrt{k'^2+m^2}}\cos\left(\sqrt{k'^2+m^2}t\right)\mathrm{e}^{-\mathrm{i}\boldsymbol{k}'\cdot\boldsymbol{x}}\right)\\=&\,m\big(\varDelta_1(x)\nabla\varDelta(x)-\varDelta(x)\nabla\varDelta_1(x)\big),\end{aligned} \tag{A-2}$$

where $\varDelta(x)$ is given by (1), $\varDelta_1(x)$ is[2]

$$\varDelta_1(x)=\int\frac{\mathrm{d}^3p}{(2\pi)^3}\mathrm{e}^{\mathrm{i}\boldsymbol{p}\cdot\boldsymbol{x}}\frac{\cos\left(\sqrt{p^2+m^2}t\right)}{\sqrt{p^2+m^2}}=\frac{1}{4\pi r}\frac{\partial}{\partial r}\begin{cases}Y_0(ms), & |\boldsymbol{x}|<t;\\ -\dfrac{2}{\pi}K_0(ms_1), & |\boldsymbol{x}|>t.\end{cases} \tag{A-3}$$

In the above formula, both $Y_0(ms)$ and $K_0(ms)$ are Bseeel functions, $s=\sqrt{t^2-\boldsymbol{x}^2}$ ,

$s_1 = \sqrt{\boldsymbol{x}^2 - t^2}$ .

From (1), when $|\boldsymbol{x}| > t$ , $\Delta(x) = 0$ , $\nabla\Delta(x) = 0$ , we have

$$\boldsymbol{J}(t, \boldsymbol{x}) = m\big(\Delta_1(x)\nabla\Delta(x) - \Delta(x)\nabla\Delta_1(x)\big) = 0\,; \qquad \text{(A-4)}$$

From (1) and (A-3), when $|\boldsymbol{x}| < t$ , we have

$$\begin{aligned}\boldsymbol{J}(t, \boldsymbol{x}) &= m\big(\Delta_1(x)\nabla\Delta(x) - \Delta(x)\nabla\Delta_1(x)\big) \\ &= m\left(\frac{1}{4\pi r}\frac{\partial Y_0(ms)}{\partial r}\nabla\left(\frac{1}{4\pi r}\frac{\partial J_0(ms)}{\partial r}\right) - \frac{1}{4\pi r}\frac{\partial J_0(ms)}{\partial r}\nabla\left(\frac{1}{4\pi r}\frac{\partial Y_0(ms)}{\partial r}\right)\right),\end{aligned} \qquad \text{(A-5)}$$

using one formula about Bseeel functions $J_\nu(z)$ and $Y_\nu(z)$ [6]:

$$\left(\frac{\mathrm{d}}{z\mathrm{d}z}\right)^m \left[z^{-\nu} Z_\nu(z)\right] = (-1)^m z^{-\nu-m} Z_{\nu+m}(z)\,,$$

where $Z_\nu(z)$ represents the Bseeel functions $J_\nu(z)$ or $Y_\nu(z)$ , (A-5) becomes

$$\boldsymbol{J}(t, \boldsymbol{x}) = \frac{m^4}{16\pi^2}\frac{\boldsymbol{x}}{s^3}\big(J_2(ms)Y_1(ms) - Y_2(ms)J_1(ms)\big),$$

and again, using another formula about $J_\nu(z)$ and $Y_\nu(z)$ :

$$J_\nu(z)Y_{\nu-1}(z) - Y_\nu(z)J_{\nu-1}(z) = \frac{2}{\pi z}\,,$$

(A-5) becomes

$$\boldsymbol{J}(t, \boldsymbol{x}) = \frac{m^4}{16\pi^2}\frac{\boldsymbol{x}}{s^3}\frac{2}{\pi m s} = \frac{1}{(2\pi)^3}\frac{m^3\boldsymbol{x}}{\left(t^2 - \boldsymbol{x}^2\right)^2}\,. \qquad \text{(A-6)}$$

Combining (A-4) and (A-6), we have

$$\boldsymbol{J}(t\,, \boldsymbol{x}) = \begin{cases} \dfrac{1}{(2\pi)^3}\dfrac{m^3\boldsymbol{x}}{\left(t^2 - \boldsymbol{x}^2\right)^2}, & |\boldsymbol{x}| < t\,; \\ 0\,, & |\boldsymbol{x}| \ge t\,. \end{cases} \qquad \text{(A-7)}$$

From (A-1) and (A-7) we obtain (54).